\documentclass[12pt]{article}

\usepackage{pstricks}
\usepackage{graphicx}
\usepackage[latin2]{inputenc}
\usepackage{epic}
\usepackage{latexsym}
\usepackage{epsf}
\usepackage{epsfig}

\newcommand{\cref}[1]{Chapter \ref{c.#1}}

\def\nn{\nonumber \\}

\def\beq{\begin{equation}}
\def\eeq{\end{equation}}
\def\bea{\begin{eqnarray}}
\def\eea{\end{eqnarray}}
\def\ba{\begin{array}}
\def\ea{\end{array}}
\def\bi{\begin{itemize}}
\def\ei{\end{itemize}}
\def\be{\begin{enumerate}}
\def\ee{\end{enumerate}}
\def\beq{\begin{equation}}
\def\eeq{\end{equation}}
\def\bc{\begin{center}}
\def\ec{\end{center}}

\def\pa{\partial}

\def\co{{\mathcal O}}

\def\ov{\overline}

\def\ds{\displaystyle}

\def\re{{\rm Re} \,}

\newcommand\Mvariable[1]{#1}
\begin{document}
\begin{flushright}
                   IFT-05-05, KAIST-TH 2005/04, DESY-05-052
\end{flushright}
\vskip 1 cm

\begin{center}
{\Huge Soft Supersymmetry Breaking in KKLT Flux Compactification}
\vspace*{1cm}
\end{center}
\noindent
\centerline{\bf K. Choi ${}^a$, A. Falkowski ${}^{b\,c}$, H. P. Nilles ${}^{d\,e}$,
M. Olechowski ${}^b$}
\vskip .7cm
\centerline{\em ${{}^{a}}$  Department of Physics, Korea Advanced Institute of Science and Technology}
\centerline{\em Daejeon 305-701, Korea}
\vskip 2mm
\centerline{\em ${{}^{b}}$ Institute of Theoretical Physics, Warsaw
  University}
\centerline{\em ul.\ Ho\.za 69, PL-00 681 Warsaw, Poland}
\vskip 2mm
\centerline{ ${{}^{c}}$ \em Deutsches Elektronen-Synchrotron DESY}
\centerline{\em Notkestrasse 85, 22607 Hamburg, Germany}
\vskip 2mm
\centerline{\em ${{}^{d}}$ Physikalisches Institut, Universit\"at Bonn}
\centerline{\em Nussallee 12, D-53115 Bonn, Germany}
\vskip 2mm
\centerline{\em ${{}^{e}}$ Theory Division, Physics Department}
\centerline{\em CERN, CH-1211 Geneva 23, Switzerland}
\vskip .5cm

\centerline{\bf Abstract}
\vskip .1cm
We examine the structure of soft supersymmetry breaking terms in KKLT models
of flux compactification with low energy supersymmetry.
Moduli are stabilized by  fluxes and nonperturbative
dynamics while a de Sitter vacuum is obtained by adding
supersymmetry breaking anti-branes.
We discuss the characteristic pattern of mass scales in such a set-up
as well as some features of 4D $N=1$ supergravity
breakdown by  anti-branes.
Anomaly mediation is found to always give an important contribution
and one can  easily arrange for flavor-independent soft terms.
In its most attractive realization, the modulus mediation is comparable to
the anomaly mediation, yielding a quite distinctive sparticle
spectrum. In addition, the axion component of the modulus/dilaton superfield
dynamically cancels the relative CP phase between the
contributions of anomaly and  modulus mediation,
thereby avoiding  dangerous SUSY CP violation.

\vskip .2cm

\thispagestyle{empty}
\setcounter{page}{0}

\newpage

\section{Introduction}

In this paper, we discuss soft terms in the KKLT scenario \cite{Kachru:2003aw} with its three steps
to achieve a supersymmetry  breaking Minkowski (or de Sitter) vacuum, while stabilizing all moduli.
This is a continuation of our previous paper \cite{Choi:2004sx} which discussed the stability and soft terms in flux compactification. The KKLT set-up has been proposed specifically within Type IIB theory
on a Calabi-Yau orientifold and we will focus on this scenario  here,
noting that  it will be straightforward to
apply our analysis to similar scenarios in other string theories \cite{Choi:2004sx}.
The first step of the procedure is to introduce the NS and RR three form fluxes \cite{fluxes},
$H_3$ and $F_3$, stabilizing \cite{Giddings:2001yu} the dilaton $S$ and all complex structure moduli $Z^\alpha$.
We shall assume the existence of  a set of flux vacua
for which the resulting imaginary self dual $G_3=H_3-i S F_3$ is aligned nearly in the direction of
a primitive $(2,1)$-form, which allows for realization of low-scale supersymmetry.
Such fluxes might fix  $m_{S,Z}$ rather close to the string or Planck scale,
while keeping the gravitino mass  much lower than $m_{S,Z}$. Thus
high scale gauge unification and weak scale supersymmetry can be realized simultaneously.
In the second step, one introduces nonperturbative dynamics, e.g. gaugino condensation
\cite{Derendinger:1985kk},
to stabilize the K\"ahler modulus $T$ at an $N=1$ supersymmetric AdS vacuum with $m_T={\cal O}(4\pi^2m_{3/2})$
and vacuum energy density $V_F\simeq -3m_{3/2}^2M_{Pl}^2$.
The third step amounts to adding  anti-D3 branes ($\bar{D3}$)
stabilized at the tip of a Klebanov-Strassler type throat \cite{Klebanov:2000hb} which has been produced by fluxes \cite{Kachru:2002gs}.
Such $\bar{D3}$ can  provide a positive contribution to the potential which would
allow the fine tuning of the total vacuum energy density to the desired
positive but small value.
It induces also a SUSY breaking vacuum shift which would result in the soft terms of visible sector fields.

In fact, the KKLT set up can be seen as  a specific example of a more general scenario (see also \cite{Brustein:2004xn})
in which (i) most of the moduli are stabilized by a high scale dynamics that leaves
$N=1$ SUSY (approximately) unbroken, (ii) a few light moduli which were
unfixed by the high scale dynamics are stabilized by nonperturbative
effects leading to an $N=1$ SUSY AdS vacuum,
(iii) this SUSY AdS vacuum is uplifted to a SUSY-breaking Minkowski (or de Sitter) vacuum by
branes which break $N=1$ SUSY explicitly.
Many of our results will be relevant for any of such scenarios.

The scheme, although quite general, leads to a specific pattern of mass scales,
as summarized in equation (\ref{massscales}). The mass scales in the low energy theory exhibit a moderate hierarchy
 characterized by the factor $\log(M_{Pl}/m_{3/2})$ that is numerically $\co (4 \pi^2)$.
The $F$-terms of the light moduli (and, in
consequence, also the soft terms in the visible sector) are suppressed
by this factor with respect to the  gravitino mass, while the moduli masses are enhanced  (by the same factor).
This leads to interesting consequences for the pattern of soft terms and the cosmological moduli problem.
In particular, when this scheme is realized in its most attractive form,
the soft terms are determined by a mixed modulus-anomaly mediation
yielding a sparticle spectrum which can be clearly distinguished
from those predicted by other mediation mechanisms.

The organization of this paper is as follows.
In the next section, we briefly review the KKLT set up, mainly focusing on the
pattern of mass scales
in those models that allow for  low energy (weak scale) supersymmetry
together with a vanishing cosmological constant.
In Section 3, we discuss soft terms in 4D $N=1$ supergravity coupled to
a SUSY breaking anti-brane, including the loop-induced
anomaly mediated contributions \cite{Randall:1998uk}
which, in the models under consideration,
turn out to be equally important as the classical moduli mediated contributions
\cite{Soni:1983rm}.
In Section 4, we examine a variety of 4D models which can be identified as
possible low energy limits of KKLT-type compactification.
We compute vevs of the auxiliary components of the moduli and chiral compensator fields
and discuss the characteristic pattern of the resulting soft terms for visible matter located on D3 or D7 branes.
Phenomenological aspects of the scheme concerning the SUSY flavor and CP problems and also the low energy sparticle spectrums will be briefly addressed.
Section 5 contains a critical summary and outlook.

\section{Scales in flux compactification}

\subsection{Mass scales in type IIB string theory}

In Type IIB  theory compactified on CY orientifold,
the 4D Planck scale $M_{Pl}$ and the 4D gauge coupling constant $g_p$ on Dp branes wrapping
$(p-3)$-cycle are given by (see e.g. \cite{po})
\begin{eqnarray}
M_{Pl}^2&=&\frac{2e^{-2\phi}}{(2\pi)^7\alpha^{\prime 4}}V_{CY}
=4\pi M^2_{st} \left(\frac{e^{-\phi}}{2\pi}\right)^2\left(M_{st}R\right)^6\, ,
\nonumber \\
\frac{1}{g_p^2}&=&\frac{e^{-\phi}}{(2\pi)^{(p-2)}(\alpha^{\prime })^{(p-3)/2}}V_{(p-3)}
=\left(\frac{e^{-\phi}}{2\pi}\right)\left(M_{st}R\right)^{(p-3)}\, ,
\end{eqnarray}
where $e^\phi=g_{st}$ and $\alpha^\prime=1/M_{st}^2$ denote the string coupling and
the string tension, respectively.
Here we use the approximation that the $(p-3)$-cycle volume is given by
$V_{(p-3)}\simeq (2\pi R)^{(p-3)}$ for
the compactification radius $R$ defined by the CY volume $V_{CY}\equiv (2\pi R)^6$.
For D7/D3 system \cite{d73}, the 4D dilaton/modulus chiral superfields $S$ and $T$  are defined as
\begin{eqnarray}
S=\frac{e^{-\phi}}{2\pi}+i c_0\, ,
\quad
T=\frac{e^{-\phi}}{2\pi}\left(M_{st}R\right)^4+i c_4\, ,
\end{eqnarray}
where $c_0$ and $c_4$ are the axions from the RR 0-form and 4-form, respectively.

With the above definitions of $S$ and $T$, the D7/D3 gauge kinetic functions
are given by
\begin{eqnarray}
f_7 = T\, ,
\qquad
f_3 = S \, .
\end{eqnarray}
Then gaugino condensation in a hidden sector on D7(D3)  generates the superpotential
\bea
W_{np}\,\sim\, e^{-8\pi^2 T/b_0}\quad
(e^{-8\pi^2 S/b_0}),
\eea
 where $b_0$ is the beta function coefficient of the gauge group in which the condensation occurs
($b_0 = N_c$ for a pure SYM theory with $SU(N_c)$ gauge group).
On the other hand,  the Euclidean D3 instanton wrapping the 4-cycle of $T$ generates
\bea
W_{np}\,\sim\, e^{-8\pi^2 T}\, .
\eea
thus the effect of the D3 instantons  is similar to that of gaugino condensation on D7.

It is convenient to express the ratios
$M_{Pl}/M_{st}$ and $M_{st}/M_{KK}$ ($M_{KK}\equiv 1/R$ is the compactificaton scale of the CY manifold) in compactified
string theory in terms of $S$ and $T$ \cite{Choi:1997an,Munoz:1999kh}.
For the Type IIB theory under consideration, one easily finds
\begin{eqnarray}
\label{expression}
\frac{M_{Pl}}{M_{st}} &=&
{4\pi ({\rm Re}(S))^{1/4}({\rm Re}(T))^{3/4}},
\nonumber \\
\frac{M_{st}}{M_{KK}}&=& \left(\frac{{\rm Re}(T)}{{\rm Re}(S)}\right)^{1/4}.
\end{eqnarray}
As we can read off from equation (\ref{expression}),  the ratios are determined
by the vacuum expectation value of the $T$ and $S$ moduli. In concrete mechanisms of moduli stabilization,
a possible range of these vevs is usually constrained. We now investigate this issue in the set-up proposed by KKLT.

\subsection{Mass scales in the KKLT scheme}

In the leading approximation,  the K\"ahler potential of the closed string moduli,
i.e. $S$, $T$ and the complex structure moduli $Z^\alpha$, is given by
\begin{eqnarray}
K=-\ln(S+S^*)-3\ln(T+T^*)-\ln i\int\Omega\wedge\Omega^*,
\end{eqnarray}
where $\Omega$ is the holomorphic $(3,0)$-form of CY.
The quantized NS and RR fluxes over 3-cycles,
\begin{eqnarray}
M_{st}^2\int_{C_3} H_3 = 4\pi^2n,\qquad
M_{st}^2 \int_{C_3} F_3 = 4\pi^2 m,
\end{eqnarray}
give rise \cite{Gukov:1999ya} to the superpotential\footnote{In this paper we use the convention that the  K\"ahler potential $K$ and the superpotential $W$ are dimensionless. These correspond to
$K/M_{Pl}^2$ and $W/M_{Pl}^3$ for the standard dimensionful K\"ahler potential and superpotential of 4D SUGRA.}
\begin{eqnarray}
\label{flux-superpotential}
W_{flux}=M_{st}^2\int G_3\wedge \Omega \, ,
\end{eqnarray}
where $G_3=F_3-iSH_3$.
Although the fluxes generate warped geometry, the warping is not significant
in most of the  region of CY except for a small region containing the Klebanov-Strassler
throat.  Then the dynamics of  bulk degrees of freedom such as the moduli
are not affected significantly by the warping.

The resulting supergravity equations of motion for $S$ and $Z^\alpha$,
$
D_Z W_{flux}= D_S W_{flux} =0
$,
generically fix $S$ and $Z^\alpha$ at a value of order unity.
The resulting $G_3$ is  imaginary self-dual on CY, $G_3=i^*G_3$, and
can be decomposed into two components:
$
\langle G_3\rangle = G_{(0,3)} + G_{(2,1)},
$
where
$
\langle W_{flux}\rangle = M_{st}^2 \int G_{(0,3)}\wedge \Omega$.
In order to have weak scale supersymmetry  {\it without} invoking to a large compactification radius
(which, as we explain later, is not available in the KKLT set up) one needs to choose a flux configuration yielding $G_3$  almost aligned  to the $(2,1)$-direction:
\begin{eqnarray}
\label{fluxhierarchy}
 \frac{G_{(0,3)}}{G_{(2,1)}} \,\ll \,1.
\end{eqnarray}
Then the resulting flux-induced gravitino mass is suppressed as
\begin{eqnarray}
(m_{3/2})_{flux} = M_{Pl} e^{K/2}W_{flux}
={\cal O}\left(\frac{e^\phi}{M_{st}^2R^3}\frac{G_{(0,3)}}{G_{(2,1)}}\right),
\end{eqnarray}
Still the masses of $Z^\alpha$ and $S$ might receive a contribution from
the unsuppressed $G_{(2,1)}$, thus
\begin{eqnarray}
m_{Z} \,\sim\, M_{Pl} e^{K/2}\frac{\partial^2_Z W_{flux}}{\partial_Z\partial_{Z^*}K}
={\cal O}\left(\frac{e^\phi}{M_{st}^2R^3}\right)\,,
\end{eqnarray}
and a similar expression holds for $m_S$ with
$\partial_Z^2 W_{flux} \to \partial_Z\partial_S W_{flux}$.

The condition (\ref{fluxhierarchy}) requires a fine tuning, where
the fluxes quantized in units of the string scale conspire to give only a tiny
supersymmetry breaking effect.
A study of the landscape of flux vacua suggests that the number of flux vacua
with small $G_{(0,3)}/G_{(2,1)}$ scales as \cite{Ashok:2003gk}
$N_{vac}(G_{(0,3)}/G_{(2,1)}\leq \epsilon)\,\sim \,\epsilon^2 N_{TOT}$,
 where $N_{TOT}$ denotes the total number of flux vacua.
Since $N_{TOT}$ can be argued to be as large as $10^{300}$ for a typical CY orientifold,
 there could still be  a large number of flux vacua for which $ G_{(0,3)}/G_{(2,1)}\sim 10^{-13}$
and hence $m_{3/2}$ is in a TeV range.  Thus, although this might not be easy to achieve in a specific model,
the KKLT set-up might be able to  accomodate the weak scale SUSY by fine-tuning the flux configuration in an appropriate manner.

The 3-form fluxes do not generate a potential for the K\"ahler modulus $T$
which corresponds to the size of a 4-cycle $C_4$.
To fix $T$, one needs to introduce additional dynamics depending on the volume of
$C_4$. As discussed earlier,  gaugino condensation on D7 branes wrapping $C_4$ and/or the Euclidean D3 instantons wrapping $C_4$
 induce a nonperturbative superpotential of the form
\begin{eqnarray}
W_{np} = C \; e^{-a \; T}\,,
\end{eqnarray}
where $C$ depends on $Z^\alpha$ in general, while $a$ is a real positive constant
of ${\cal O}(4\pi^2)$ (in our normalization of $T$).

For the flux configuration
with $G_{(0,3)}/G_{(2,1)}\sim 10^{-13}$ yielding $m_{Z,S}={\cal O}(M_{Pl})$ and $m_{3/2}
={\cal O}(1)$ TeV, the stabilization of $T$ and also the SUSY breaking
can be described by an effective SUGRA theory which is obtained after integrating out
 $S$ and $Z^\alpha$. This is tacitly assumed in the original set-up of KKLT \cite{Kachru:2003aw}.
It requires
a certain decoupling between the heavy fields $Z, S$ on one side and the light modulus
$T$ on the other. In some cases (e.g. when $ \partial_Z\partial_S W_{flux} \ll \partial_Z^2 W_{flux}$), such a decoupling might not be possible \cite{Choi:2004sx} and a second modulus (e.g. $S$) might be light and remain in the
low-energy effective theory. We shall later give examples of both of these
situations, but stick to the simplest case in the following discussion.

The effective $N=1$ SUGRA thus  contains the $T$-modulus as well as the SM gauge
and chiral matter fields originating from D7/D3 branes.
Although the whole set-up includes also SUSY breaking $\bar{D3}$ branes,
the VEV of $T$ is mainly determined by the $N=1$ SUSY sector described by
the following effective K\"ahler and superpotential:
\begin{eqnarray}
\label{effective}
K_0&=&-3\ln(T+T^*),
\nonumber \\
W_0&=&w_0 -  Ce^{-aT},
\end{eqnarray}
where $w_0=\langle W_{flux}\rangle ={\cal O}(G_{(0,3)}/G_{(2,1)})$ and $C={\cal O}(1)$
are constants in the effective theory.
This effective theory successfully stabilizes $T$ as it has a unique SUSY AdS vacuum yielding
\begin{eqnarray}
\label{tstabilization}
\langle  a  \,{\re T} \rangle &\simeq& \ln (M_{st}/m_{3/2})={\cal O}(4\pi^2),
\nonumber \\
 m_{3/2} &=& \langle M_{Pl} e^K W_0\rangle
\simeq {M_{Pl} w_0 \over (2 \re T)^{3/2}},
\nonumber \\
\langle V_{N=1}\rangle &=&-3m_{3/2}^2 M_{Pl}^2 \, ,
\end{eqnarray}
where $V_{N=1}$ is the standard $N=1$ SUGRA potential for the K\"ahler and superpotential of (\ref{effective}).
Note that in the KKLT set-up  the large weak/string  scale hierarchy results in
the appearance of a moderately large parameter
 $a  \,{\re T}$. This parameter will enter the expressions for SUSY breaking order parameters,
leading to a moderate hierararchy between various soft masses. Another observation is that, since
$a={\cal O}(1) \div {\cal O}(4 \pi^2)$, we also obtain  $\re T =
{\cal O}(4 \pi^2)\div {\cal O}(1)$. Larger values of $\re T$ are not available within this stabilization  scheme.
In consequence, the resulting values of $g_{st}$ and $M_{st} R$ are rather close to unity,
and both $M_{st}$ and $M_{KK}$ are {\it not} far away from the Planck scale.

As we have remarked, for a gaugino condensation from $SU(N_c)$ SYM theory,
$a={8\pi^2}/{N_c}$ and thus $\ln(M_{st}/m_{3/2})\approx {8\pi^2}/{N_c g_7^2}$,
where $g_7^2$ is the 4D gauge coupling on D7  at the compactification scale.
If the SM gauge fields originate from D7,
this relation amounts to  $w_0\sim e^{-16\pi^2/N_c}$, implying that
in this case the flux fine tuning for a hierarchically small $w_0\sim
G_{(0,3)}/G_{(2,1)}$ is necessary also to get the correct value of the SM gauge coupling,
unless $N_c$ is unusually
large as ${\cal O}(8\pi^2)$.

To obtain a phenomenologically desirable de Sitter or Minkowski vacuum, KKLT proposed
to introduce a $\bar{D3}$ providing a positive potential energy.
In the presence of 3-form fluxes the geometry is warped, which can be
parameterized  as \cite{Giddings:2001yu}
\begin{eqnarray}
ds^2=e^{2A(y)} ({\rm Re}(T))^{-3/2}g^{E}_{\mu\nu}dx^\mu dx^\nu+e^{-2A(y)}
({\rm Re}(T))^{1/2}\tilde{g}_{mn}dy^mdy^n,
\end{eqnarray}
where $g^E_{\mu\nu}$ is the 4D Einstein frame metric and
$\tilde{g}_{mn}$ is a CY metric normalized as
$\int d^6y \sqrt{\tilde{g}}=M_{st}^{-6}$.
Here we have ignored the fluctuation of $S$ and $Z^\alpha$
in the 10D metric since $m_{Z,S}={\cal O}(M_{st})$.
As long as the flux density is not strong,
the warp factor $e^{A(y)}$ would be of order unity over the most region of CY, however
it can be exponentially small around the small region of Klebanov-Strassler (KS) throat.
It is then a plausible assumption that the SM lives on D7/D3 which are stabilized in the region
where the warping is not significant.
On the other hand, $\bar{D3}$ favors to be stabilized at the tip of the KS throat, $y=y_{\bar{D3}}$,
where the warp factor
is minimal and exponentially small \cite{Giddings:2001yu}:
\begin{eqnarray}
e^{A(y_{\bar{D3}})}=e^{A_{min}}\sim ({\rm Re}(T))^{1/4}e^{-2\pi n/g_{st}m} \, ,
\end{eqnarray}
where $n=\frac{1}{4\pi^2}M_{st}^2\int_{C_{KS}}H_3$ and $m=\frac{1}{4\pi^2}M_{st}^2
\int_{\tilde{C}_{KS}}F_3$ are integers for the NS and RR fluxes over
the collapsing 3-cycle $C_{KS}$ and its dual $\tilde{C}_{KS}$ of the KS throat.
Adding the $\bar{D3}$ tension to the negative vacuum energy density $V_{N=1}\simeq -3m_{3/2}M_{Pl}^2$
induced by gaugino condensation,
the total vacuum energy density would be given by
$V_{TOT} \approx V_{N=1}+V_{\bar{D3}}$.
Given this we see that it might be possible
to adjust the observed value of the cosmological constant by a careful
fine tuning of the parameters.
A simple calculation of the $\bar{D3}$ tension in the 4D Einstein frame
gives
\begin{eqnarray}
\label{antid3-up}
V_{\bar{D3}} \sim \frac{(e^{A_{min}}M_{Pl})^4}{[{\rm Re}(T)]^3}
\sim  \frac{(e^{-2\pi n/g_{st}m}M_{Pl})^4}{[{\rm Re}(T)]^2}.
\end{eqnarray}
Then the condition $V_{TOT}\simeq 0$ requires that
\begin{eqnarray}
e^{A_{min}}\sim \sqrt{m_{3/2}/M_{Pl}} \, .
\end{eqnarray}
With this warping the physical cutoff scale on $\bar{D3}$ will
be of the order of the intermediate scale $e^{A_{min}}M_{st}\sim
\sqrt{m_{3/2}M_{Pl}}$ and the $\bar{D3}$ moduli
acquire  masses of ${\cal O}(e^{A_{min}}/M_{st}^2R^3)$, also close
to the intermediate scale.

As we see, only a limited class of flux vacua can give
weak scale SUSY together with vanishing cosmological constant:
those with flux configurations with $G_{(0,3)}/G_{(2,1)}\sim 10^{-13}$
 and the warping  $e^{A_{min}}\sim  \sqrt{G_{(0,3)}/G_{(2,1)}}\sim 3\times 10^{-7}$.
Such two step fine tunings might be a generic feature of
any realistic model which incorporates the weak scale Higgs mass and the small cosmological
constant within the string landscape whose typical mass scale is not far below $M_{Pl}$.

Below we summarize all the mass scales in KKLT set-up which realizes  weak scale SUSY
together with a vanishing vacuum energy density through the effective
SUGRA (\ref{effective}) and the uplifting potential (\ref{antid3-up}):
\begin{eqnarray}
\label{massscales}
M_{st} &\sim& 5\times 10^{17} \,\, {\rm GeV},\nonumber \\
1/R &\sim& 10^{17} \,\, {\rm GeV}, \nonumber \\
m_{Z,S} &\sim& \frac{1}{M_{st}^2R^3} \,\sim\, 10^{16} \,\, {\rm GeV}\, ,
\nonumber \\
\Lambda_{GC}&=&M_{st}e^{-\langle aT\rangle/3} \,\sim\, 10^{13}\,\, {\rm GeV} \, ,
\nonumber \\
M_{\bar{D3}} &\sim& e^{A_{min}}M_{st} \,\sim\,10^{11} \,\, {\rm GeV}\, ,
\nonumber \\
m_T &\sim& \langle a T\rangle m_{3/2} \,\sim\, 10^5 \,\, {\rm GeV} \, ,
\nonumber \\
m_{3/2} &\sim& \frac{1}{M_{st}^2R^3}\left(\frac{G_{(0,3)}}{G_{(2,1)}}\right)
\,\sim\, 10^4 \,\, {\rm GeV}\, ,
\nonumber \\
m_{soft} &\sim& M_{weak}\sim \frac{m_{3/2}}{\langle aT\rangle}
\,\sim\, 10^2 \div 10^3  \,\, {\rm GeV}\, .
\end{eqnarray}
Here $\Lambda_{GC}$ is the dynamical scale of D7 gaugino condensation,
$M_{\bar{D3}}$ is the red-shifted cutoff scale on $\bar{D3}$,
The expressions for $m_T$ and $m_{soft}$ will be derived in Section 3.
There we shall also consider examples with more than one light modulus.

\subsection{Comments on alternative schemes}

So far we have considered a scenario with a TeV scale gravitino mass. Besides weak scale supersymmetry
the consequence of this approach is that both the string scale $M_{st}$ and compactification scale $M_{KK}$ are close to the the Planck scale (so that the standard high scale gauge unification is possible).
An alternative \cite{Burgess:1998px} is to consider the fundamental string scale in Type IIB theory at an intermediate value $M_{st}\sim 10^{10}$ GeV. From eq. (\ref{expression}), this would require a stabilization of $S$ and $T$ at values satisfying
\begin{eqnarray}
 {\rm Re}(S){\rm Re}(T)^{3}\,\sim\, 10^{30} \, .
\end{eqnarray}
This is difficult within the known mechanisms of moduli stabilization (but see \cite{Balasubramanian:2005zx})  and is certainly not possible within the KKLT stabilization scheme. But even if that could be achieved there is another apparent problem.  For the SUGRA approximation to be valid
one  needs  $M_{KK}\leq M_{st}$, which by   eq. (\ref{expression}) requires  ${\rm Re}(T)\,\geq \,{\rm Re}(S)$.
Then  $g_7^2 = {\rm Re}(T)^{-1} \,\leq \,3\times 10^{-8}$, which means that  the SM cannot be embedded in the D7 sector.
One might avoid this conclusion by assuming that a 4 cycle has a volume $V_4\sim M_{st}^{-4}$, so $g_7^2 \sim 1$, while its dual 2 cycle has a huge volume
$V_2\sim 10^{15} M_{st}^{-2}$.

Another scheme might involve placing the D3 and/or D7 branes at a warped
throat \cite{DeWolfe:2002nn} and explain the hierarchy between the
weak scale and the Planck scale by a strong warping instead of weak scale
supersymmetry. We shall not discuss these schemes here, as we are
mainly concerned with the question of soft SUSY breaking terms.


One might also be interested in alternatives to the ad-hoc uplifting procedure of KKLT.
In Ref. \cite{Burgess:2003ic}, it has been proposed to uplift the AdS vacuum to a dS or Minkowski vacuum
by introducing gauge field flux on a D7 brane.
Such two-form flux would induce a moduli-dependent FI term and thus a positive $D$-term potential.
A virtue of this scenario would be that the uplifting can be described within
the standard 4D $N=1$ SUGRA framework, thus the resulting SUSY breaking
can be interpreted as a spontaneous breaking.

To examine the possibility of $D$-term uplifting,
let us briefly review the $F$ and $D$-term scalar potentials in generic 4D SUGRA \cite{Binetruy:2004hh}.
The scalar potential in the Einstein frame is
$V=V_F+V_D$
where
\begin{eqnarray}
V_F&=& M_{Pl}^4 e^K\left(K^{I\bar{J}}D_I W D_{\bar{J}}W^*- 3|W|^2\right),
\nonumber \\
V_D&=&\frac{1}{2} M_{Pl}^4 {\rm Re}(f_a) D^a D^a
= \frac{1}{2 {\rm Re}(f_a)} M_{Pl}^4 \left(i \eta^I_a \partial_I K - 3 i r_a\right)^2 \, .
\end{eqnarray}
Here $D_I W =\partial_I W +\partial_IK W$ is the K\"ahler covariant derivative of the superpotential.
Furthermore,  $\eta_a^I$ denotes the gauge transformation of chiral superfields under the gauge group factor $G_a$,
 $\delta_a \Phi^I = \eta_a^I(\Phi)$,
while  $r_a$ is determined by the transformation properties of
the superpotential under $G_a$,
\beq
\label{gaugetransform}
\delta_a W =\eta_a^I\partial_I W =-3 r_a W \, .
\eeq
 This formalism can accomodate both a field dependent FI term  accompanying  the Green-Schwarz mechanism
(when $\eta^S = i \delta_{GS}$ and $\pa_S K = -(S+ \ov S)^{-1}$)
as well as a field independent FI term occuring in the presence of gauged $R$-symmetry (for $r_a = {\rm const} \neq 0$).
But from (\ref{gaugetransform}) for $W\neq 0$ one easily finds that the D-terms can be rewritten as
\beq
D^a = \frac{i}{{\rm Re}(f_a)}\frac{1}{W}\eta_a^I D_I W \, .
\eeq
Obviously, an arbitrary supersymmetric configuration with $\langle D_I W\rangle=0$,
if allowed, would be a stationary solution of $V_F$.
It is clear that if $V_F$ admits a supersymmetric AdS vacuum, i.e. a stable solution
with $\langle D_I W\rangle =0$ but $\langle W\rangle\neq 0$, as in the KKLT case,
this SUSY AdS solution remains a good vacuum solution of
the complete potential $V=V_F+V_D$. So in this case uplifting with a
D-term will not work.
If the absolute minimum is at $\langle D_I W\rangle=0$,
$V_D$ improves the stability of the SUSY AdS vacuum of $V_F$ without lifting
the vacuum energy density.
Thus, if one wishes to get a (SUSY breaking) dS or Minkowski vacuum
within the standard 4D SUGRA, one needs to have a special form of $W$ which
does not admit a  SUSY AdS solution.
It remains an open question whether a viable model of this kind can be derived from string theory
(see \cite{Jockers:2005zy} for a recent study).

\section{Soft terms}

\subsection{N=1 supergravity coupled to anti-D3 brane}

An intriguing feature of the KKLT scenario is
the final step to introduce $\bar{D3}$ which uplifts the SUSY AdS vacuum
to a SUSY breaking Minkowski (or de Sitter) vacuum.
The $N=1$ SUSY preserved by the combined dynamics of flux and gaugino condensation
is not respected on the worldvolume of $\bar{D3}$, in particular there
are no light open string degrees of freedom  on $\bar{D3}$ which
would form a $N=1$ supermultiplet.
In this subsection, we discuss the 4D effective action describing the effects of $\bar{D3}$
on the low energy dynamics of the light moduli and the SM fields in KKLT-type
flux compactification.

Our main problem is to find the low energy couplings between
the $N=0$ sector on $\bar{D3}$ and the $N=1$ sector propagating in the 10D bulk or the D7/D3
worldvolumes.
A full description of the couplings between a supersymmetric bulk sector and
a less supersymmetric brane sector usually requires  an off-shell formulation of
the bulk supersymmetry.
In our case, such couplings can be described most conveniently by spurion operators
in $N=1$ superspace.
Since D7/D3 branes do not intersect with $\bar{D3}$,
spurion operators on the worldvolume of $\bar{D3}$ do not depend on
the fields confined in D7/D3, but generically depend on the bulk fields
as well as on the fields confined in $\bar{D3}$.
It is rather generic that all scalar fields on $\bar{D3}$ get a mass of
the order of $e^{A_{min}} M_{st}\sim 10^{10}$ GeV by the flux, thus are heavy enough to
be decoupled at TeV scale.
There may exist some light fields with non-zero spin on $\bar{D3}$ but those will not affect our analysis of the soft terms.
This allows us to focus on
the $\bar{D3}$ spurion operators depending only on the light bulk moduli and 4D SUGRA multiplet.
As we will see, in such situation the dominant effects of $\bar{D3}$ on the other parts of
the theory can be described by a single spurion operator which corresponds to
the $\bar{D3}$ tension.

Let us consider the 10D action at energies below $M_{st}$  of the following  form:
\begin{eqnarray}
S_{10D}=\int d^{10}x \left[
{\cal L}_{IIB}+\delta^2(z-z_{D7}){\cal L}_{D7}+
\delta^6(y-y_{D3}){\cal L}_{D3}+\delta^6(y-y_{\bar{D3}}){\cal L}_{\bar{D3}}\right],
\end{eqnarray}
where ${\cal L}_{IIB}$ is the 10D IIB SUGRA action,
${\cal L}_{Dp}$ denotes the worldvolume action of Dp brane, and
$z$ and $y$ correspond to the transverse coordinates of D7 and D3 branes, respectively.
The 4D effective action of the light fields is of the form
$
S_{4D}=S_{N=1}+S_{\bar{D3}},
$
where $S_{N=1}$ is the $N=1$ SUGRA action originating from
${\cal L}_{IIB}$ and ${\cal L}_{D7/D3}$, while $S_{\bar{D3}}$
is from the non-supersymmetric ${\cal L}_{\bar{D3}}$.
The $N=1$ supersymmetric part can be written in the standard
superspace form:
\begin{eqnarray}
\label{N=1}
S_{N=1}&=&\int d^4x \sqrt{g^C} \,\left[\,
\int d^4\theta \,
CC^*\left(-3\exp(-K_{eff}/3)\right)
\right.\nonumber \\
&&+\,\left.\left\{
\int d^2\theta
\left(\frac{1}{4}f_a W^{a\alpha}W^a_\alpha
+C^3W_{eff}\right)
+{\rm h.c.}\right\}\,\right],
\end{eqnarray}
where  $g^C_{\mu\nu}=(CC^*)^{-1}e^{K_{eff}/3}g^E_{\mu\nu}$ for the 4D Einstein-frame
metric $g^E_{\mu\nu}$ and the chiral compensator superfield
$C=C_0+\theta^2 F^C,$
and the K\"ahler and superpotential can be expanded as
\begin{eqnarray}
-3\exp[-K_{eff}/3]&=&
-3\exp[-K_0(\Phi^m,\Phi^{m*})/3]+Y_i(\Phi^m,\Phi^{m*})Q_i^*Q_i,
\nonumber \\
W_{eff}&=& W_0(\Phi^m)+\frac{1}{6}\lambda_{ijk}(\Phi^m)Q_iQ_jQ_l,
\nonumber \\
f_a&=& f_a(\Phi^m),
\end{eqnarray}
where $\Phi^m$ denote the gauge singlet light  moduli and $Q_i$ stand for
the gauge charged matter fields.
In (\ref{N=1}), we have ignored the 4D SUGRA multiplet other than the metric component.
In this scheme, $C_0$ is
a redundant degree of freedom, which is reflected by  the invariance under
the following Weyl transformation which is
a part of the super Weyl invariance in the full compensator formulation:
\begin{eqnarray}
\label{superweyl}
C\,\rightarrow\, e^{-2\tau} C,
\quad
g_{\mu\nu}\,\rightarrow\, e^{2(\tau+\bar{\tau})}g_{\mu\nu},
\quad
\theta^\alpha \rightarrow e^{-\tau+2\bar{\tau}}\theta^\alpha,
\end{eqnarray}
where $\tau$ is a complex constant.

As stressed, $S_{\bar{D3}}$ does not depend on the matter and gauge superfields confined in D7/D3.
Then at the leading order in the supercovariant derivative expansion, $S_{\bar{D3}}$ can be
written as
\begin{eqnarray}
\label{N=0}
S_{\bar{D3}}&=&\int d^4x \sqrt{g^C} \int d^4\theta \,
\left[\,-\frac{1}{2}e^{4A_{min}}C^2C^{2*}\theta^2
\bar{\theta}^2P(\Phi^m,\Phi^{m*})\right.\nonumber \\
&&\left.\qquad +\,
e^{3A_{min}}C^3\bar{\theta^2}\Gamma(\Phi^m,\Phi^{m*})
+{\rm h.c.}\,\right],
\end{eqnarray}
where $P$ and $\Gamma$ are model-dependent functions of $\Phi^m$ and $\Phi^{m*}$,
which are generically of order unity (in units with $M_{Pl}=1$), and
$e^{A_{min}}$ is  the (generically  moduli-dependent) warp factor on  $\bar{D3}$.
The $C$-dependence of $S_{\bar{D3}}$ can be determined by requiring invariance of the $\bar{D3}$ action under the Weyl transformation (\ref{superweyl}). Then the power of warp factor in each spurion operator is determined
by the $C$-dependence since both $C_0$ and the warp factor correspond
to the conformal mode of the 4D metric.

The $N=1$ SUSY appears to be explicitly broken in (\ref{N=0}).
In fact, the $N=1$ SUSY might be non-lineraly realized as  suggested in \cite{bagger}
through a Goldstino fermion $\xi^\alpha$ living on the worldvolume of $\bar{D3}$.
In this case, the $\bar{D3}$ action (\ref{N=0}) can be
extended  to a manifestly supersymmetric form
by replacing the $D$-spurion $\theta^2\bar\theta^2$
by a real superfield $\Lambda^2\bar\Lambda^2$,
and the $F$-spurion by a chiral superfield
$(\bar{\cal D}^2-8{\cal R})\Lambda^2\bar\Lambda^2$,
where $\Lambda^\alpha$ is the Goldstino superfield defined in \cite{Samuel:1982uh}:
\bea
\Lambda^\alpha=\theta^\alpha+\frac{1}{M^2_{\bar{D3}}}\xi^\alpha+...
\eea
for $M_{\bar{D3}}\sim e^{A_{min}}M_{st}$,
and $(\bar{\cal D}^2-8{\cal R})$ is the chiral projection operator of
4D SUGRA. Then the spurion operators of (\ref{N=0}) can be obtained from the
following form of super-Weyl invariant action involving the Goldstino superfield:
\bea
S_{\bar{D3}}=\int d^4x \sqrt{g^C}
\int d^4\theta \,\tilde{C}\tilde{C}^*\,{\cal L}_{\bar{D3}}(\frac{\tilde{C}^{1/2}}{\tilde{C}^*}{\cal D}_\alpha,
\frac{\tilde{C}^*}{\tilde{C}^{1/2}}\Lambda^\alpha, \Phi^m) \, ,
\eea
where $\tilde{C}=e^{A_{min}}C$.

The coefficient function $e^{4A_{min}}P$ can be easily computed
for the minimal KKLT model in which the only light modulus is
the overall volume modulus $T$.  In such model, the warp factor
on $\bar{D3}$ depends on $T$ as  $e^{A_{min}}=\left({\rm Re}(T)\right)^{1/4}e^{-2\pi n/g_{st}m}$
\cite{Giddings:2001yu}.
Matching the $\bar{D3}$ tension to (\ref{N=0}) under
the relation between the 4D Einstein frame metric $g^E_{\mu\nu}$ and the
string frame metric $g_{s\mu\nu}=e^{2A(y)}\left({\rm Re}(T)\right)^{-3/2}g^E_{\mu\nu}$,
one finds
\begin{eqnarray}
e^{4A_{min}}e^{2K_0/3}P= \frac{D}{(T+T^*)^2} \, ,
\end{eqnarray}
where $D$ is a constant of order $M_{Pl}^4 e^{-8\pi n/g_{st}m}$.

\subsection{Soft terms in the presence of anti-D3 brane}

In the presence of $S_{\bar{D3}}$ which breaks SUSY explicitly or
realizes SUSY non-linearly, the resulting soft terms are modified
compared to the known results in standard 4D SUGRA.
Solving the equations of motion for the auxiliary fields in $S=S_{N=1}+S_{\bar{D3}}$,
one easily finds
\begin{eqnarray}
\label{auxiliary}
\frac{F^C}{C_0}&=&\frac{1}{3}\partial_IK_{eff}F^I+\frac{C_0^{*2}}{C_0}e^{K_{eff}/3}\left(
W_{eff}+e^{3A_{min}}\Gamma\right)^*,
\nonumber \\
F^I&=&-\frac{C_0^{*2}}{C_0}e^{K_{eff}/3}K_{eff}^{I\bar{J}}\left(\,D_J
\left(W_{eff}+e^{3A_{min}}\Gamma\right)\,\right)^*,
\end{eqnarray}
where $\{ \Phi^I\}=\{\Phi^m, Q_i\}$, the K\"ahler covariant derivative
$D_IX=\partial_IX+(\partial_IK_{eff})X$, and one can choose $C_0=e^{K_{eff}/6}$ to arrive at the
Einstein metric frame.
It is then straightforward to find the following moduli potential
and the soft parameters of the canonically normalized visible fields in the Einstein frame:
\begin{eqnarray}
\label{potential}
V_0&=&e^{K_0}\left[\,K_0^{m\bar{n}}D_m(W_0+e^{3A_{min}}\Gamma)\left(D_n(W_0+e^{3A_{min}}\Gamma)\right)^*\right.
\nonumber \\
&&\left.\qquad -3|W_0+e^{3A_{min}}\Gamma|^2\,\right]+ e^{4A_{min}}e^{2K_0/3}P,
\nonumber \\
{\cal L}_{soft}&=& -m_i^2|\tilde{Q}_i|^2 
- \left (\frac{1}{2}M_a\lambda^a\lambda^a
+\frac{1}{6}A_{ijk}y_{ijk}\tilde{Q}_i\tilde{Q}_j\tilde{Q}_k+{\rm h.c.} 
\right ) \, ,
\end{eqnarray}
where
\begin{eqnarray}
\label{soft}
M_a&=& F^m\partial_m\ln\left({\rm Re}(f_a)\right),
\nonumber \\
A_{ijk}&=& -F^m\partial_m\ln\left(\frac{\lambda_{ijk}}{Y_iY_jY_k}\right),
\nonumber \\
&=& -F^m\left(\,\partial_m K_0 +\partial_m\ln\left(\frac{\lambda_{ijk}}{Z_iZ_jZ_k}\right)\,\right),
\nonumber \\
m_i^2 &=& \frac{2}{3}V_0-F^mF^{n*}\partial_m\partial_{\bar{n}}
\ln \left(Y_i\right)
\nonumber \\
&=& -\frac{1}{3}e^{4A_{min}}e^{2K_0/3}P+
\left(V_0+m_{3/2}^2-F^mF^{n*}\partial_m\partial_{\bar{n}}\ln \left(Z_i\right)\right)
\end{eqnarray}
for the canonically normalized Yukawa couplings
\begin{eqnarray}
y_{ijk}=\frac{\lambda_{ijk}}{\sqrt{Y_iY_jY_k}}.
\end{eqnarray}
Here $Z_i=e^{K_0/3}Y_i$ is the K\"ahler metric of $Q_i$, i.e.
\begin{eqnarray}
K_{eff}=K_0(\Phi^m,\Phi^{m*})+Z_i(\Phi^m,\Phi^{m*})Q_iQ_i^*,
\end{eqnarray}
and $m_{3/2}$ is the gravitino mass containing both the standard $N=1$ contribution from
$W_0$ and the contribution from $\bar{D3}$:
\begin{eqnarray}
m_{3/2}=  M_{Pl} e^{K_0/2} \left(W_0+e^{3A_{min}}\Gamma\right).
\end{eqnarray}
In the literature \cite{Soni:1983rm},
$A_{ijk}$ and $m_i^2$ are normally expressed in terms of
the moduli K\"ahler potential $K_0$ and the matter K\"ahler metric $Z_i$.
Our results (\ref{soft}) show that it is more convenient to express those soft parameters
in terms of the superspace wavefunction coefficient $Y_i=e^{-K_0/3}Z_i$,
particularly when the effects of the SUSY breaking $\bar{D3}$ are included.

Obviously,  when $P=\Gamma=0$,  (\ref{auxiliary}), (\ref{potential}) and (\ref{soft})
become the standard expressions of the SUSY breaking auxiliary components,
the moduli potential and the soft terms in $N=1$ SUGRA.
In KKLT models, in the absence of $\bar{D3}$, $W_0$ leads to a SUSY AdS vacuum.
Generically $P$ and $\Gamma$ are of order one
for the moduli VEV of order unity, while $W_0$ is of order $m_{3/2}/M_{Pl}$.
Then, as we have noticed in sec. 2.2,
in order for the AdS vacuum to be uplifted to a Minkowski vacuum
by $\bar{D3}$, one needs
$e^{A_{min}}\,\sim\, \sqrt{m_{3/2}/M_{Pl}}$.
This implies that
\begin{eqnarray}
e^{3A_{min}}\Gamma \,\sim\, \sqrt{m_{3/2}/M_{Pl}}\,W_0,
\end{eqnarray}
and thus the contribution to $m_{3/2}$ from $e^{3A_{min}}\Gamma$
is negligible compared to the contribution from $W_0$ induced
by the flux and gaugino condensations.
This means that $e^{3A_{min}}\Gamma$ can be safely ignored,
so the effects of $\bar{D3}$ on the low energy dynamics of
the moduli and visible fields can be described well by a single coefficient function $P$.
In such case, the SUSY breaking auxiliary components are well approximated by the
standard $N=1$ expressions:
\begin{eqnarray}
\label{approx-F}
\frac{F^C}{C_0}&\simeq &\frac{1}{3}\partial_m K_0 F^m+ M_{Pl}e^{K_0/2} W_0^*,
\nonumber \\
F^m&\simeq &-  M_{Pl} e^{K_0/2} K_0^{m\bar{n}}\left(D_n
W_0\right)^*,
\nonumber \\
m_{3/2}&\simeq & M_{Pl} e^{K_0/2}W_0\, ,
\end{eqnarray}
and the moduli scalar potential is approximately given by
\begin{eqnarray}
\label{approx-potential}
V_0\simeq M_{Pl}^4 e^{K_0}\left(K^{m\bar{n}}_0 D_m W_0 (D_nW_0)^*-3|W_0|^2\right)+V_{\rm lift} \, ,
\end{eqnarray}
where the uplifting potential from $\bar{D3}$ is given by
\begin{eqnarray}
V_{\rm lift}=e^{4A_{min}}e^{2K_0/3}P \, .
\end{eqnarray}
Thus once $K_0$ and $W_0$ for the $N=1$ sector and $P$ for ${\bar{D3}}$
are given, one can compute the SUSY breaking order parameters $F^m,F^C$ and $m_{3/2}$
using the above approximate results \cite{Choi:2004sx}.

The soft parameters in (\ref{soft}) correspond to the moduli-mediated
tree level contributions at the compactification scale, and do not
include possible loop effects.
As we will see, in KKLT-type models, $F^C/C_0\simeq m_{3/2}$ is
bigger than the moduli-mediated soft masses typically by a factor
$\sim 4\pi^2$.
With this little hierachy, the anomaly-mediated
contributions  from $F^C/C_0$ \cite{Randall:1998uk}
will be equally important as the moduli-mediated ones although they involve
the loop suppression factor $1/8\pi^2$.
Then the dominant parts of the soft terms at energies just below the compactification scale
can be obtained by simply adding the anomaly-mediated and also anomaly-moduli-mixed
contributions to the moduli-mediated results (\ref{soft}).
As is well known, such loop contributions are determined by the RG runnings
of $Y_i$  and $f_a$, which are given by
\bea
\Delta {\rm Re}(f_a)&=& -\frac{1}{16\pi^2}\left(3T_a({\rm Adj})-\sum_i T_a(Q_i)\right)\ln\left(\frac{
CC^*}{\mu^2}\right)\, ,
\nonumber \\
\Delta \ln\left(Y_i\right)&=&-\frac{1}{32\pi^2}\left(
4\sum_a\frac{C_a(Q_i)}{{\rm Re}(f_a)}-\sum_{jk}\frac{|\lambda_{ijk}|^2}{Y_iY_jY_k}
\right)\ln\left(\frac{CC^*}{\mu^2}\right)\, ,
\nonumber \eea
where $T_a$ denotes the quadratic Casimir
and $\mu$ is the renormalization point.
Then replacing $f_a$ and $\ln(Y_i)$ in (\ref{soft}) by
$f_a+\Delta f_a$ and $\ln(Y_i)+\Delta \ln(Y_i)$, and also $F^m\partial_m$ by
$F^m\partial_m+F^C\partial_C$, we find the soft masses just below the compactification scale
\begin{eqnarray}
\label{soft1}
M_a&=& F^m\partial_m\ln\left({\rm Re}(f_a)\right) +\frac{b_ag_a^2}{8\pi^2}\frac{F^C}{C_0}\, ,
\nonumber \\
A_{ijk}&=&
-F^m\partial_m\ln\left(\frac{\lambda_{ijk}}{Y_iY_jY_k}\right)-
\frac{\gamma_i+\gamma_j+\gamma_k}{16\pi^2}\frac{F^C}{C_0}\, ,
\nonumber \\
m_i^2&=& \frac{2}{3}V_0-F^mF^{n*}\partial_m\partial_{\bar{n}}\ln \left(Y_i\right)
-\frac{1}{32\pi^2}\frac{d\gamma_i}{d\ln\mu}\left|\frac{F^C}{C_0}\right|^2
\nonumber \\
&+&\frac{1}{16\pi^2}\left({\gamma}_{m}^i
F^m\left(\frac{F^C}{C_0}\right)^*
+{\rm h.c}\right)\, ,
\end{eqnarray}
where $b_a=-\frac{3}{2}T_a({\rm Adj})+\frac{1}{2}\sum_i T_a(Q_i)$ are the one-loop beta function coefficients,
$\frac{dg_a}{d\ln\mu}=\frac{b_a}{8\pi^2} g_a^3$,
$\gamma_i=2\sum_a  g_a^2 C_a(Q_i)-\frac{1}{2}\sum_{jk}|y_{ijk}|^2$ are
the anomalous dimension of $Q_i$,
$\frac{1}{8\pi^2}\gamma_i=\frac{d\ln Y_i}{d\ln\mu}$,
$T_a$ and $C_a$ are group theory factors, 
${\rm Tr} (T^2) = T_a(Q_i)$, $\sum T^2 = C_a(Q_i) I$,   
and finally ${\gamma}_m^i=\partial_m\gamma_i$ are given by
\begin{eqnarray}
{\gamma}_{m}^i
=-\frac{1}{2}\sum_{jk}|y_{ijk}|^2\partial_m\ln\left(\frac{\lambda_{ijk}}{Y_iY_jY_k}\right)
-2\sum_a g_a^2 C_a(Q_i)\partial_m\ln\left({\rm Re}(f_a)\right).
\end{eqnarray}
If there is a Higgs bilinear term
$H_uH_d$ either in the superpotential or in the K\"ahler potential,
there will arise a soft $B$-term with $B\sim m_{3/2}$ at tree level.
Since $m_{3/2}\sim 4\pi^2 M_a$ in our case, such $B$ would be too large to
be phenomenologically acceptable.  A simple way to avoid this difficulty is to
assume that the Higgs $\mu$-term originates from a trilinear Yukawa term $\lambda NH_uH_d$
with an additional singlet field $N$,
which would give $\mu=\lambda\langle N\rangle$ and $B=A_{NH_uH_d}$.

\section{Models}

In the previous section, we discussed the soft terms for a general form of
4D effective action $S_{4D}=S_{N=1}+S_{\bar{D3}}$ which  describes the low energy
limit of  KKLT flux compactification.
In this section, we consider some specific examples and
compute the SUSY breaking order parameters, i.e. $F^C$ and $F^\Phi$ ($\Phi=$ moduli),
as well as the resulting soft terms of visible fields.

We will examine two simple classes of models,
one in which  $T$ is the only light modulus,
and the other in which both $T$ and $S$ appear as light moduli.
The first class of models represents a typical situation in Type IIB flux
compactification with low energy SUSY, in which  the dilaton $S$ and all complex structure moduli
$Z^\alpha$ acquire  superheavy masses $m_{S,Z}\sim 1/M_{st}^2R^3\sim 10^{16}$ GeV
from fluxes, while the K\"ahler modulus $T$ gets  $m_T\sim m_{3/2} \ln(M_{Pl}/m_{3/2}) \sim 10^6$ GeV from gaugino condensation.
However, as was pointed out in \cite{Choi:2004sx}, $S$ can be light also
in some cases. For instance, if the NS fluxes are all vanishing, the flux-induced
superpotential (\ref{flux-superpotential}) would be independent of $S$ and $T$,
 and then only $Z^\alpha$ get  superheavy masses of ${\cal O}(1/M_{st}^2R^3)$
from the RR fluxes.
In this class of models there is no supersymmetric minimum before the uplifting.
This runaway behavior can be cured  if the superpotential acquires both $S$ and $T$ dependence  from
 non-perturbative dynamics, for instance by
gaugino condensations on $D3$  and $D7$ branes yielding
$W_{np}=Ce^{-aT}+\Lambda e^{-bS}$.
As for the uplifting potential $V_{\rm lift}=e^{4A_{min}}e^{2K_0/3}P$ (in the Einstein
frame) from $\bar{D3}$, we will use $D/t^{n_t}$ for the first class of models, and
$D/t^{n_t}s^{n_s}$ for the second class of models, where
$T=t+i\tau$, $S=s+i\sigma$ and $D$ is a constant.
As we will see, in all KKLT type models
the anomaly mediation is generically equally or sometimes even more
important than the moduli  mediation.

\subsection{Models with light K\"ahler moduli}

The simplest  KKLT-type model is the original one \cite{Kachru:2003aw}
described by
\begin{eqnarray}
\label{model1_moduli}
\mbox{Model  1}: \quad K_0&=&-3\ln(T+T^*),
\nonumber \\
W_0&=& w_0 - C_1 e^{-aT}\, ,
\nonumber \\
V_{\rm lift}&=&D/t^{n_t} \, ,
\quad
(T=t+i\tau) \, .
\end{eqnarray}
For $K_0$ depending only on $t$, the overall phase of $W_0$ is irrelevant, and
also the relative phase between  $w_0$
and $C_1$ can be eliminated by shifting the axion field $\tau$ without
 any physical consequence.
Then we can choose both
$\omega_0$ and $C_1 $ to be real and positive without loss of generality,
in  which case $\langle \tau\rangle=0$.
Using the scalar potential given by (\ref{approx-potential})
and also the expressions  (\ref{approx-F}) for SUSY breaking order parameters,
we find the model has a stable SUSY breaking Minkowski vacuum (when $D$ is  tuned
to make $\langle V_0\rangle =0$) with the following vevs and masses:
\begin{eqnarray}
\frac{F^C}{C_0}\,\approx\,m_{3/2}&\approx& M_{Pl} \frac{w_0}{2\sqrt{2}t^{3/2}}\, ,
\nonumber \\
\frac{F^T}{(T+T^*)}&\approx&\frac{n_t}{2 a\,t} m_{3/2}\, ,
\nonumber \\
m_t\,\approx\, m_{\tau} &\approx& 2 a\,t\,m_{3/2}\, ,
\end{eqnarray}
where
\bea
at \approx \ln (C_1/w_0) \sim 4\pi^2\, ,
\eea
and $n_t=2$ if $V_{\rm lift}$ is from $\bar{D3}$.
Note that the parameter $C_1$ is of the order of $1$, while
$w_0$ is of the order of $m_{3/2}/M_{Pl}$, and then
$a t$ is of order $4\pi^2$ for $m_{3/2}$  close to the TeV scale.
Such value of $at$ gives rise to a little hierarchy between the moduli mass $m_T$, the gravitino
mass and the gaugino/sfermion soft masses:
\bea
m_T\sim 4\pi^2 m_{3/2}\sim (4\pi^2)^2 m_{soft} \, .
\eea
A particularly interesting feature of the model  is that
\bea
\frac{F^T}{(T+T^*)}\sim \frac{1}{4\pi^2}\frac{F^C}{C_0}\, ,
\eea
and as a consequence
the anomaly-mediation always gives a non-negligible
contribution to soft masses.

In some cases, fluxes might preserve accidently a (discrete) $R$-symmetry,
and thereby yield $w_0=0$. Still $T$ can be stabilized by introducing multiple
gaugino condensations. Such case can be described by
\begin{eqnarray}
\mbox{Model 2}:\quad  K_0&=&-3\ln (T+T^*),
\nonumber \\
W_0&=&C_1e^{-a_1T}-C_2e^{-a_2T},
\nonumber \\
V_{\rm lift}&=&D/t^{n_t} \quad
(T=t+i\tau),
\end{eqnarray}
where we can choose $C_1$ and $C_2$ to be real and positive without loss of generality and then set $\langle\tau\rangle =0$.
In order to stabilize $t$ at a value yielding hierarchically
small $m_{3/2}/M_{Pl}$,
one needs to tune $a_{1,2}$ as
\bea
|a_1-a_2|\approx \frac{a_1+a_2}{\ln(M_{Pl}/m_{3/2})} \, ,
\eea
as in the standard racetrack model \cite{Krasnikov:1987jj}.
We then find
\bea
\ds \frac{F^C}{C_0}\,\approx \,m_{3/2} &\approx&\ds
\frac{a_2-a_1}
  {2\,{\sqrt{2}}\,a_1t^{\frac{3}{2}}}
M_{Pl} C_2e^{-a_2t},
\nn
\ds
{F_T \over (T+T^*)} &\approx& \ds \frac{3\,\Mvariable{n_t}}{4\,\Mvariable{a_1}t\,\Mvariable{a_2}\,t}  m_{3/2},
\nn
\ds
m_t \,\approx\, m_\tau  &\approx&
\frac{4 a_1\,t\,a_2\,t}{3} m_{3/2},
\eea
where
\bea
a_1 t\approx a_2 t \approx \ln (M_{Pl}/m_{3/2}) \, .
\eea
The most important feature of this model is that $\frac{F_T}{(T+T^*)} \sim \frac{m_{3/2}}{(a \, t)^2}$ where
$at\approx  4\pi^2$, and as a consequence the soft terms are dominated
by the anomaly-mediated contributions.

\subsection{Models with light dilaton and K\"ahler moduli}

In this subsection, we examine some models in which both $T$ and $S$ are stabilized
by a nonperturbative superpotential.
Let us first consider the model given by
\begin{eqnarray}
\mbox{Model 3}:\quad
K_0&=&-\ln(S+S^*)-3\ln(T+T^*) \, ,
\nonumber \\
W_0&=&w_0-C_1 e^{-aT}-\Lambda e^{-bS}\, ,
\nonumber \\
V_{\rm lift}&=&D/t^{n_t}s^{n_s} \quad
(T=t+i\tau, \,S=s+i\sigma)\, ,
\end{eqnarray}
where $C_1 e^{-aT}$ is induced  by the gaugino condensation on $D7$,
while $\Lambda e^{-bS}$ is from the gaugino condensations on $D3$.
Again we can choose $w_0$, $C_1 $ and $\Lambda$ to be real and positive
with  appropriate shifts of $\tau$ and $\sigma$, and then
$\langle \tau\rangle=\langle\sigma\rangle=0$.
We then find
\bea
at&\approx& \ln (C_1 /w_0)\approx \ln (M_{Pl}/m_{3/2})\, ,
\nonumber \\
bs&\approx& \ln (\Lambda/w_0)\approx \ln(M_{Pl}/m_{3/2})\, ,
\eea
and also
\bea
\frac{F^C}{C_0}\,\approx\, m_{3/2}
&\approx& \frac{w_0}{4s^{1/2}t^{3/2}}\, ,
\nonumber \\
\frac{F^T}{(T+T^*)}&\approx & \frac{n_t}{2at} m_{3/2}\, ,
\nonumber \\
\frac{F^S}{(S+S^*)}&\approx &  \frac{3n_s}{2bs}m_{3/2} \, .
\eea
together with the dilaton/moduli masses
\bea
m_{S,T}\approx m_{3/2}\ln(M_{Pl}/m_{3/2})\, .
\eea

Again the flux-induced $w_0$ in $W_0$ can be accidently vanishing,
and $T$ and $S$ can be  stabilized by having multi gaugino condensations
either on $D7$ or on $D3$.
In case that two gaugino condensations arise from $D7$ branes, one would have
\bea
\mbox{Model 4:}\quad
K_0&=&-3\ln(T+T^*)-\ln(S+S^*)\, ,
\nonumber \\
W_0&=&C_1e^{-a_1T}-C_2e^{-a_2T}-\Lambda e^{-bS}\, ,
\nonumber \\
V_{\rm lift}&=&D/t^{n_t}s^{n_s}\, .
\eea
For this model, we find
\bea
a_1t\approx a_2t\approx bs\approx \ln(M_{Pl}/m_{3/2})\, ,
\eea
and
\bea
\frac{F^C}{C_0}\,\approx\, m_{3/2}&\approx& \frac{bs^{1/2}}{2t^{3/2}}\frac{\Lambda e^{-bs}}{M_{Pl}^2}\, ,
\nonumber \\
\frac{F^T}{(T+T^*)}&\approx& \frac{3n_t}{4a_1t\,a_2t}m_{3/2}\, ,
\nonumber \\
\frac{F^S}{(S+S^*)}&\approx& \frac{3n_s}{2bs}m_{3/2}\, .
\eea

The above results show that $\frac{F^S}{(S+S^*)}\sim \frac{F^C}{4\pi^2 C_0}$, thus
$F^S$ and $F^C$ can give  equally important contributions to soft terms.
On the other hand, the $F$-component of $T$ stabilized by the racetrack superpotential
is further suppressed as $\frac{F^T}{(T+T^*)}\sim \frac{F^C}{(4\pi^2)^2C_0}$, thus
gives negligible contribution to soft terms.
Similarly, one can consider a model with
\bea
W_0=\Lambda_1 e^{-b_1S}-\Lambda_2 e^{-b_2 S}-Ce^{-aT},
\eea
and then one finds
$\frac{F^T}{(T+T^*)}\sim \frac{F^C}{4\pi^2 C_0}$ and
$\frac{F^S}{(S+S^*)}\sim \frac{F^C}{(4\pi^2)^2 C_0}$.

\subsection{Some phenomenological features}

Let us briefly discuss some phenomenological features of the
soft terms in the models considered above.
For the models of sec. 4.1, the matter K\"ahler metric and the holomorphic
Yukawa couplings and gauge kinetic functions would be
given by
\bea
\label{model1_matter}
\Delta K_{eff}&=&\frac{1}{(T+T^*)^{n_i}}Q^*_iQ_i\, ,
\nonumber \\
\Delta W_{eff}&=& \frac{1}{6}\lambda_{ijk}Q_iQ_jQ_k\, ,
\nonumber \\
f_a &=& T^{\,l_a}\, ,
\eea
where $\lambda_{ijk}$ are constants,
$l_a=1$, $n_i=0$ and $l_a=0$, $n_i=1$ for gauge and matter fields living on D7 and D3, respectively.
In the case when  matter fields live on intersections of $D7$ branes
$n_i$ take a value between 0 and 1 \cite{Ibanez:2004iv}.
Using the general result (\ref{soft1}), the soft parameters of visible fields
at energies just below the compactification scale are found to be \cite{Choi:2004sx}
\bea
\label{soft_model1}
M_a &=&l_a\frac{F^T}{(T+T^*)}+\frac{b_ag_a^2}{2}\left(\frac{F^C}{4\pi^2 C_0}\right),
\nonumber \\
A_{ijk}&=&(3-n_i-n_j-n_k)\frac{F^T}{(T+T^*)}
-\frac{1}{4}(\gamma_i+\gamma_j+\gamma_k)\left(\frac{F^C}{4\pi^2 C_0}\right)\, ,
\nonumber \\
m_i^2&=&(1-n_i)\left|\frac{F^T}{(T+T^*)}\right|^2
-\frac{1}{32\pi^2}\frac{d\gamma_i}{d\ln\mu}\left|\frac{F^C}{C_0}\right|^2
\nonumber \\
&&+\left(
\frac{1}{8}\sum_{jk}(3-n_i-n_j-n_k)|y_{ijk}|^2 -
\frac{1}{2}\sum_a l_aT_a(Q_i)g_a^2\right)
\nonumber \\
&& \times\left(\frac{F^T}{(T+T^*)}\left(\frac{F^{*C}}{4\pi^2 C_0^*}\right)
+\frac{F^{*T}}{(T+T^*)}\left(\frac{F^{C}}{4\pi^2 C_0}\right)\right) \, .
\eea

For the models with the visible sector living on D3,
we have $l_a=0$ and $n_i=1$, so $F_T$ does not contribute to the soft terms at
tree-level. The no-scale structure is lifted by loop corrections and dependence of the soft terms on $F_T$ will appear at one-loop level.
But, since $\frac{F^T}{(T+T^*)}\sim \frac{F^C}{4\pi^2C_0}$ or even smaller,
 the soft terms are dominated by anomaly mediation.
As is well known, pure anomaly mediation leads to negative slepton masses squared.
In view of this point, the most attractive possibility would be  the visible sector living on D7,
 in particular the model 1
in which the modulus mediation and the anomaly mediation give comparable contributions to the soft terms
(for the model 2 $F_T$ is too small to cure the negative slepton masses problem).
It turns out \cite{Okumura} that
the low energy phenomenology of such mixed modulus-anomaly mediation
is quite different from
the pure anomaly mediation \cite{Randall:1998uk}
and also from the pure modulus mediation \cite{Allanach:2005yq}.
In particular, since $F^T$ and $F^C$ have the same sign,
the anomaly mediation cancels the RG evolution
of the modulus-mediation, leading to a quite distinctive
superparticle spectrum at low energy scale \cite{Okumura}.

It should be stressed that the soft terms (\ref{soft_model1}) have been derived
within a framework  satisfying two important conditions:
(i) all relevant moduli are stabilized and (ii) the vacuum energy density is correctly tuned to be nearly zero.
Unless the condition (i) is met, additional dynamics  should be introduced to stabilize the unfixed moduli, and this new dynamics might change the predictions completely.
The condition (ii) is also important for  reliable computation of soft scalar masses
since any additional source of vacuum energy density generically affects the soft scalar masses
\cite{Choi:1994xg}.
To our knowledge, this is the first example to compute soft terms in string theory framework
satisfying these  two conditions simultaneously.

To examine the structure of soft terms in models with light $T$ and $S$, one can consider
the following forms of the matter K\"ahler metric and the
gauge kinetic functions:
\bea
\Delta K_{eff}&=&\frac{1}{(S+S^*)^{k_i}(T+T^*)^{n_i}}Q^*_iQ_i \, ,
\nonumber \\
\Delta W &=& \frac{1}{6}\lambda_{ijk}Q_iQ_jQ_k \, ,
\nonumber \\
f_a &=& \kappa_a S+l_aT \, ,
\eea
where $(n_i,k_i)=(0,1)$ and $(n_i,k_i)=(1,0)$ for matter fields on $D7$ and $D3$, respectively,
and $(\kappa_a,l_a)=(0,1)$ and $(\kappa_a,l_a)=(1,0)$ for gauge fields living on $D7$ and $D3$, respectively.
The resulting soft terms can be easily obtained from (\ref{soft1})
as in \cite{Choi:2004sx}.
 In this case a viable superparticle spectrum is obtained also for the visible sector living on D3.

Independently of the detailed low energy phenomenology \cite{Okumura},
the soft parameters predicted by KKLT set-up have an attractive feature avoiding
naturally the SUSY flavor and CP problems.
The soft terms  preserve the quark and lepton flavors if $n_i$, or $(n_i,k_i)$ for the models
with light $T$ and $S$, are flavor-independent,
which would arise automatically if the matter fields with common
gauge charges live on the same $D$-brane worldvolume (or their intersection).
They also preserve CP since the relative CP phase between
$F^T$ and $F^C/C_0$ could be rotated away by the shift of the axion-like
field $\tau$.
Such dynamical relaxation of the potentially dangerous SUSY CP phase
can be considered to be a consequence of an approximate nonlinear PQ symmetry
$T\rightarrow T+i\alpha$ ($\alpha=$ constant) which is broken
by the stabilizing superpotential $W_0$ \cite{Choi:1993yd}.
A similar relaxation of SUSY CP phases can be achieved in more general cases
with multi-moduli $\Phi^m$ if the following conditions are met:
(i) $W_0=\sum_p C_p\exp (-\sum_m a_m^p\Phi^m)$ and $f_a =
\sum_m l_{am}\Phi^m$ for real parameters  $a^p_m$ and $l_{am}$,
(ii) the moduli K\"ahler potential,
matter K\"ahler metric, and the uplifting potential depend only on $\Phi^m+\Phi^{*m}$,
(iii) the number of independent terms in $W_0$ is limited as
$N_{W_0}\leq N_\Phi+1$ where $N_\Phi$ is the number of involved moduli.
In fact, these three conditions are satisfied in all
models examined in this section.
Finally we remark that the little mass hierarchy driven by
the factor $\ln(M_{Pl}/m_{3/2})\sim 4\pi^2$, i.e.
 $m_T\sim 4\pi^2 m_{3/2}$ and $m_{3/2}\sim 4\pi^2 m_{soft}$, allows
the model to be free from the cosmological moduli problem and possibly also
from the cosmological gravitino problem.
It offers also an interesting scenario to produce a correct amount of neutralino
dark matter as has been studied recently \cite{Kohri:2005ru}.

\section{Conclusion}

A combination of fluxes and non-perturbative effects might allow for stabilization of
all moduli of string theory. The KKLT set-up provides a specific suggestion how this can
be achieved explicitely in the framework of the type IIB theory. It involves three
steps: (i) a breakdown of supersymmetry through fluxes, (ii) a restoration of
supersymmetry (in AdS) via nonperturbative effects, followed by (iii) again
a breakdown of supersymmetry in the process of uplifting the vacuum energy to
the desired value. The scheme requires a severe fine tuning of the fluxes to get
a weak scale supersymmetry in addition to the other fine tuning for
small cosmological constant.
One might hope
that future research could shed some light into the question of fine tuning
and offer a more elegant description of the problem.

The study of the phenomenological properties of the scheme requires a
careful analysis of the soft supersymmetry breaking terms. It is quite easy to compute the soft terms
after the first step of the KKLT procedure, but this does not lead to
meaningful results as in the second step supersymmetry is restored. From
our analysis we can actually draw a useful lesson: it does not make sense
to compute the soft terms in a scheme that has not yet stabilized all
the moduli. The stabilization of the last modulus might (and usually does)
change the results completely.
So the first condition for a reliable computation of soft terms is to stabilize
all the relevant moduli.
Another important condition is that the framework should allow
the vacuum energy density to be fine tuned to the desired small value
since any additional source of vacuum energy density generically affects
the soft scalar masses.
To our knowledge, our analysis is the first attempt to compute soft terms in string theory framework
satisfying these  two conditions simultaneously.

The result of our analysis turns out to be rather simple and appealing.
As we have shown, the KKLT scheme yields  a unique and characteristic pattern of mass
scales and soft terms. In case of a single light modulus (stabilized by nonperturbative
effects) the result on mass scales is given in equation (\ref{massscales}). We see that
apart from the large hierarchy between the Planck scale and the weak scale,
there appears a little hierarchy characterized by a numerical factor
${\cal O}(4\pi^2)$ originating from $\ln(M_{Pl}/m_{3/2})$.
One has $m_T={\cal O}(4\pi^2m_{3/2})$ as well as
$m_{3/2}={\cal O}(4\pi^2m_{soft})$. The appearance of the soft terms is
due to a specific mixed modulus-anomaly mediation, yielding the results
of (\ref{soft1}) and (\ref{soft_model1}) which give
a quite distinctive superparticle spectrum \cite{Okumura}.

Independent of the details of the low energy phenomenology the soft parameters have the
attractive feature of avoiding naturally the SUSY flavour and CP problems.
In addition, the rather large value of the gravitino mass and the mass of the
$T$-modulus could lead to interesting cosmological consequences
\cite{Kohri:2005ru}. In the case of two light moduli the
overall picture remains unchanged:
the appearance
of these little hierachies persists and
the pattern of mass scales is quite similar.

\vspace{5mm}
\noindent{\large\bf Acknowledgments}
\vspace{5mm}

This work was partially supported by the EU 6th Framework Program
MRTN-CT-2004-503369 ``Quest for Unification'' and
MRTN-CT-2004-005104 ``ForcesUniverse''.
K.C. is supported by Korean KRF PBRG 2002-070-C00022
and the Center for High Energy Physics of Kyungbook National University.
A.F. was partially supported by the Polish KBN grant 2 P03B 129 24
for years 2003-2005.
M.O. was partially supported  by the Polish KBN grant 2 P03B 001 25
for years 2003-2005.

K.C. would like to thank the theory group of Bonn University
for the hospitality during his visit.
The  stay of A.F. at DESY is possible due to Research Fellowship  granted by  Alexander von Humboldt Foundation.

\end{document}